\documentclass[twocolumn,amssymb, amsmath,nobibnotes, aps, pra, floatfix,superscriptaddress,showpacs]{revtex4-1}
\usepackage{bm}%
\usepackage{graphicx}
\usepackage[toc,page]{appendix}
\usepackage{perpage}
\usepackage{amsmath,amssymb,mathrsfs}
\usepackage{latexsym}
\usepackage{graphicx} 
\usepackage{epstopdf}
\usepackage{graphicx,epstopdf,color}
\usepackage{amsfonts}
\usepackage{amsmath,amssymb,mathrsfs}
\usepackage{hyperref}
\usepackage{bm}

\allowdisplaybreaks[1]

\begin{document}

\author{Lo\"{i}c Henriet}
\affiliation{Centre de Physique Th\'{e}orique, Ecole Polytechnique, CNRS, 91128 Palaiseau Cedex France}

\author{Zoran Ristivojevic}
\affiliation{Centre de Physique Th\'{e}orique, Ecole Polytechnique, CNRS, 91128 Palaiseau Cedex France}

\author{Peter P. Orth}
\affiliation{Institute for Theory of Condensed Matter, Karlsruhe Institute of Technology (KIT), 76131 Karlsruhe, Germany}

\author{Karyn Le Hur}
\affiliation{Centre de Physique Th\'{e}orique, Ecole Polytechnique, CNRS, 91128 Palaiseau Cedex France}

\title{Quantum Dynamics of the Driven and Dissipative Rabi Model}
\begin{abstract}
The Rabi model considers a two-level system (or spin-1/2) coupled to a quantized harmonic oscillator and describes the simplest interaction between matter and light. The recent experimental progress in solid-state circuit quantum electrodynamics has engendered theoretical efforts to quantitatively describe the mathematical and physical aspects of the light-matter interaction beyond the rotating wave approximation. We develop a stochastic Schr\"{o}dinger equation approach which enables us to access the strong-coupling limit of the Rabi model and study the effects of dissipation, and AC drive in an exact manner. We include the effect of ohmic noise on the non-Markovian spin dynamics resulting in Kondo-type correlations, as well as cavity losses. We compute the time evolution of spin variables in various conditions. As a consideration for future work, we discuss the possibility to reach a steady state with one polariton in realistic experimental conditions.
\end{abstract}

\maketitle

\section{Introduction} Originally, the Rabi model had been introduced to describe the effect of a weak and rapidly rotating magnetic field on an atom possessing a nuclear spin \cite{rabi1,rabi2}. Nowadays, this model is applied to a variety of quantum systems, from quantum optics to condensed matter physics. A few examples include microwave and optical cavity quantum electrodynamics (QED) \cite{cohen,raimond,haroche}, ion traps \cite{Wineland}, quantum dots, and superconducting qubits in circuit QED \cite{dot,dot2,Vion,alexandre,wallraff,rob,Mooij,Hofheinz,Esteve,Todorov,Niemczyk,Irfan,Martinis,DenisBernard,nicolas}. Recent on-chip experiments, by using artificial two-level systems made of superconducting qubits, allow a high control on the coupling between the system and the light field \cite{Vion,alexandre,wallraff,rob,Mooij,Hofheinz,Esteve,Todorov,Niemczyk}. Effective photon-photon interactions and photon blockade effects may also be engineered \cite{Verger,kimble,photon_blockade_1,photon_blockade_2,photon_blockade_3,bloch}. Such spin-boson systems are of importance for applications in quantum computing \cite{zollerquant, Grangier, Yale}. 

The application of the rotating wave approximation (RWA) is justified in the weak coupling limit and results in the Jaynes-Cummings (JC) model \cite{JC}, which is exactly solvable. Analytical solutions of the quantum Rabi model beyond the RWA have been recently explored based on the underlying discrete $\mathbb{Z}_2$ (parity) symmetry \cite{braak,moroz,zhong,gritsev}. Moreover, some dynamical properties of the model have been addressed \cite{braak2} and other theoretical efforts in the strong-coupling limit are achieved \cite{larson,nataf,marco,Simone}. In this article, we study the Rabi model in a wide regime of parameters, from the weak to the strong coupling, and account for external driving and non-Markovian dissipation effects on the two-level system from the environment. The latter is modeled by a bath of harmonic oscillators \cite{Caldeira} and gives rise to ohmic dissipation on the spin dynamics \cite{Leggett,weiss}. At low temperatures, this engenders a renormalized (many-body) Rabi frequency for the two-level system and non-trivial damping processes which can be measured in cold atom, ion trap, mesoscopic, and photon systems \cite{recati,peterivankaryn,sortais,dominik,dominik2,Cirac,Saleur1997,Camalet,KLH1,Moshe,peropadre,delsing,Serge,whitney,Sougato}. By introducing two stochastic fields, we extend the non-perturbative Schr\"{o}dinger equation method of Refs. \cite{stochastic,2010stoch,adilet,Lesovik,Stockburger}. We show the applicability of the stochastic method  by focusing on the spin dynamics in various conditions. We complement our results via physical and analytical arguments. We also discuss non-trivial dynamical final states with one polariton achieved by driving the system. By increasing the drive amplitude we decrease the characteristic time to reach a pure state with one polariton. This may find applications to realize a driven Mott state of polaritons, {\it i.e.}, dressed states (eigenstates) of light and matter \cite{houck}, in the weak-coupling limit between light and matter. 

\subsection{Model} 
The Hamiltonian describing the driven and dissipative quantum Rabi model reads
\begin{eqnarray}
 H_{sys} &=&\frac{\Delta}{2} \sigma^z+ \omega_0 \left(a^{\dagger}a+\frac{1}{2}\right)+\frac{g}{2} \sigma^x (a+a^{\dagger}) \\ \nonumber
 &+& V(t) (a+a^{\dagger}) + \sum_k \left[\omega_k b_k^{\dagger} b_k+ \lambda_k (b_k+b_k^{\dagger})\frac{\sigma^x}{2}\right],
\label{model}
\end{eqnarray}
where $a^{\dagger}$ and $a$ are creation and annihilation operators for the quantized harmonic oscillator with frequency $\omega_0$, $\sigma^i$ ($i=x,y,z$) are the Pauli spin operators for the spin-1/2, $\Delta$ is the resonant frequency between the two levels, and $g$ denotes the interaction strength (we set $\hbar=1$). The first line of the Hamiltonian represents the Rabi model. The term containing $V(t) = V_0\cos(\omega_d t)$ incorporates the effect of the coherent semi-classical external drive on the cavity \cite{bishop,bishop2}. Dissipation is taken into account via $b_k^{\dagger}$ and $b_k$ which are the creation and annihilation operators for the bosonic mode $k$, with frequency $\omega_k$, and $\lambda_k$ describes the microscopic interaction of the two-level system with the  environment, which we assume to be of ohmic type. The Jaynes-Cummings weak-coupling limit of the Rabi model is reproduced when neglecting the counter-rotating terms, $(\sigma^+ a^{\dagger} + h.c.)$ where $\sigma^{\pm}=(\sigma^x\pm i\sigma^y)/2$, which ensures a continuous $U(1)$ symmetry and an associated conserved quantity, the polariton number $N = a^{\dagger} a + \sigma^+ \sigma^-$. 

The combined effect of the cavity and of the ohmic bath on the spin is encapsulated through the spectral function \cite{Leggett,weiss}:
\begin{equation}
J(\omega)=\pi g^2 \delta(\omega-\omega_0)+2\pi \alpha \omega \exp(-\omega/\omega_c). 
\end{equation}
Here, $\alpha$ determines the effective (dimensionless) coupling between the spin and the bath, while $\omega_c$ is a high-frequency cutoff, which we take to be the largest energy scale in the system. The quantities $\alpha$ and $\omega_c$ are related to the parameters in the Hamiltonian via $\pi \sum_k |\lambda_k|^2 \delta(\omega - \omega_k) = 2 \pi \alpha \omega e^{- \omega/\omega_c}$.

A typical example of experimental setup is a Cooper pair box system at resonance \cite{makhlin,Clerk} where, within our notations, the operator $\sigma^x$ represents the presence or absence of excess Cooper pairs in the island. The transverse field  $\sigma^z$ can be realized by coupling the Cooper pair box to a macroscopic superconductor via the Josephson effect. We assume that the Cooper pair box is capacitively coupled to the electromagnetic cavity and that ohmic dissipation embodies resistive effects stemming from the mesoscopic circuit \cite{makhlin,Clerk,schon}. Other superconducting circuits known as flux \cite{Lloyd} or phase qubits \cite{martinis} provide equivalent systems. Superconducting systems \cite{transmon,paik,fluxonium} yield a long decoherence time which  corresponds to very small values of $\alpha$.  A similar Hamiltonian could be derived in the case of a dissipative flux qubit \cite{Astafiev}. Note, the case of very strong qubit dissipation could also be addressed both theoretically and experimentally  \cite{stochastic,saclay,glebf,FPierre,JE,Nazarov,pothier,karyn,ines,Matveev,KarynMeirong,Meirong,zarand,SergePascal,philippe,zoran}. 

\subsection{Dissipation effects}

In the absence of the cavity ($\omega_0=g=V=0$), the physics of the model (\ref{model}) is already quite rich as it describes, e.g., the ohmic spin-boson model, Kondo physics, and long-range Ising models \cite{Anderson,Guinea,Leggett,weiss}. Several methods have been devised to address the spin dynamics for the spin-boson model, such as the Non-Interacting Blip approximation (NIBA) \cite{weiss,Leggett}. The result for ${\cal P}(t)=\langle \sigma^x(t) \rangle$ can be found using Heisenberg equations of motion. It is first convenient to perform a polaronic transformation 
\begin{equation}
U=\exp\left[-\sigma^x \sum_k \frac{\lambda_k}{2\omega_k} (b^{\dagger}_k-b_k) \right],
\end{equation}
 in order to remove the spin-bath coupling term. We can then reach the NIBA equation for the dynamics of ${\cal P}(t)$ by averaging in a weak coupling sense the spin equations of motion, leading to \cite{Karyn}:
\begin{eqnarray}
\dot{{\cal P}}(t)+\int_{t_0}^t \mathcal{F}(t-t') {\cal P}(t') dt'=0,
\label{NIBA}
\end{eqnarray}
where $\mathcal{F}(t)=\Delta^2 \cos\left[ Q_1 (t)\right] \exp\left[-Q_2(t)\right]$, and:
\begin{align}
 Q_1(t)&=\int_0^{\infty} d\omega\frac{J(\omega)}{\omega^2}\sin \omega t \notag \\
 Q_2(t)&=\int_0^{\infty} d\omega\frac{J(\omega)}{\omega^2}\left(1-\cos \omega t\right) \coth \frac{\beta \omega}{2} .
\end{align}
Eq. (\ref{NIBA}) can be solved using Laplace transform, and for $0<\alpha<1/2$ the spin dynamics shows coherent damped oscillations ${\cal P}(t)=a\cos(\zeta t +\phi)\exp(-\gamma t)$, with a universal quality factor: 

\begin{align}
\frac{\zeta}{\gamma}=\cot \left( \frac{\pi \alpha}{2(1-\alpha)} \right).
\end{align}
In this calculation it appears that the energy splitting $\Delta$ between the two levels is dressed by the bosonic modes, leading to a many-body renormalization to 
\begin{equation}
\Delta_r=\Delta (\Delta/\omega_c)^{\alpha/(1-\alpha)}.
\end{equation}

Dissipative effects may also stem from the photonic part of the system. Photon leakage out of the system can be taken into account by adding an imaginary component $\Gamma$ to the photon frequency $\omega_0$. The cavity is indeed exposed to the vaccuum noise of the surrounding environment, and energy can leak out into the external bath. This effect can be addressed in a microscopic manner by considering a coupling of the inner photonic modes to an infinite number of external bosonic modes, so that the Hamiltonian becomes (within a rotating-wave approximation) \cite{Clerk2}:
\begin{align}
H=H_{sys}+\sum_q \omega_q l^{\dagger}_q l_q-i \sum_q \left[ f_q a^{\dagger}l_q +f_q^* a l^{\dagger}_q  \right],
\end{align}
where $l^{\dagger}_q$ ($l_q$) is the creation (annihilation) operator of an external boson of frequency $\omega_q$. The use of the Heisenberg equations of motion in the Markov approximation, which assumes that the coupling strength $f=\sqrt{|f_q|}$ and the density of state $\rho=\sum_q \delta(\omega-\omega_q)$ are constant, allows to write the effect of the environment as a imaginary component $\Gamma=2 \pi f^2 \rho$ for the photon frequency \cite{Clerk2,hybrid5}. 

In the following, we will mainly focus on the high-Q cavity limit where we assume that dissipation effects are more important on the two-level system rather than on the cavity, which corresponds to $\Gamma \ll \gamma$.  There exists a variety of other schemes to study the dissipative dynamics \cite{sassetti} of photonic systems, such as phenomenological Linblad \cite{Lindblad}, or Bloch-Redfield master equations \cite{Bloch,Redfield} derived from the parameters of a microscopic model. The stochastic method under consideration \cite{stochastic,2010stoch,adilet,Lesovik,Stockburger} allows to compute the out-of-equilibrium non-Markovian spin dynamics by taking into account the effect of the cavity mode, dissipation and drive in a exact manner. Note that other stochastic approaches were developed \cite{diosi,dalibard,zoller}. 

The paper is organized as follows. In Sec. II, we will derive most of the results concerning the stochastic method, based on the Feynman-Vernon influence functional approach \cite{FV} and the Blip-Sojourn approach \cite{Leggett}. In Sec. III, we will show results on the Rabi model including the drive effects. In particular, within our approach we reproduce the Bloch-Siegert shift and the strong-coupling adiabatic limit. In Sec. IV, we shall discuss dissipation, drive and lattice effects.

\section{Spin dynamics from path integral approach}

We first consider the case $\Gamma=0$ (ideal cavity) and $V(t)=0$ and we will reach an effective stochastic Schr\"{o}dinger Equation for the spin-reduced density matrix after a stochastic decoupling. 
We will then be able to compute the spin variables $\langle \sigma^x (t) \rangle$ and $\langle \sigma^z (t) \rangle$, and various initial conditions for the spin will be considered. We will finally incorporate the effects of external drive on the system.

We assume without lack of generality that the spin and bath are uncoupled at the initial time $t_{0}$ when they are brought into contact, and therefore the total density matrix can be factorized \cite{Leggett}: $\rho_{tot} (t_{0})=\rho_{B}(t_{0}) \otimes \rho_{S} (t_{0})$. Here, $\rho_{B}$ and $\rho_{S}$ are respectively the bosonic and spin reduced density matrices. In the following, we parametrize the spin path according to its value along the $x$-axis (corresponding to the direction of the coupling with the bosonic bath). We therefore choose notations in which the density matrix corresponding to a pure state along the $x$-axis is: 
 \begin{equation} \rho_{|+_x\rangle \langle +_x|}=\left( \begin{array}{cc}
1&0 \\
0&0
\end{array} \right).\end{equation}

Elements of the reduced density matrix can be expressed as:
\begin{equation}
\langle \sigma_f | \rho_S (t) | \sigma_f'\rangle= \sum_{\sigma_0,\sigma_0'} \langle \sigma_0 | \rho_{S} (t_0) |\sigma_0' \rangle \int D\sigma D\sigma' \mathcal{A}_{\sigma} \mathcal{A}_{\sigma'}^* F_{\sigma, \sigma'}.
\label{eq:densitymatrixelement}
\end{equation}
$D\sigma$ and $D\sigma'$ denote integration over all real-time spin paths $\sigma$ and $\sigma'$ with fixed initial conditions $|\sigma_0\rangle$ and $|\sigma_0'\rangle$  and final conditions $|\sigma_f\rangle$ and $|\sigma_f'\rangle$. The terms $\mathcal{A}_{\sigma}$ and $\mathcal{A}_{\sigma'}$ denote the free amplitude for the spin to follow a given path. 

The influence of the bosonic bath (photons and bosonic modes describing the dissipation) is fully contained in the Feynman Vernon influence functional $F_{\sigma, \sigma'}$ which reads \cite{FV}:
\begin{align}
F_{\sigma, \sigma'}=\exp\Big\{-\frac{1}{\pi} \int_{t_0}^t ds \int_{t_0}^s ds' \big[&-i L_1(s-s') \xi(s)\eta(s') \notag\\
&+L_2(s-s')\xi(s)\xi(s')  \big] \Big\},
\label{eq:influence}
\end{align}
where $\eta$ and $\xi$ are the symmetric and antisymmetric spin paths: $\eta (s)=\frac{1}{2} \left[\sigma(s)+\sigma'(s) \right]$, $\xi (s)=\frac{1}{2} \left[\sigma(s)-\sigma'(s) \right]$. Here the two functions $L_1$ and $L_2$ characterize the interaction with the bath:
\begin{align} &L_1(t)=\int_0^{\infty} d \omega J(\omega) \sin \omega t  \notag \\
&L_2(t)=\int_0^{\infty} d \omega J(\omega) \cos \omega t \coth \frac{\beta \omega}{2} .
\label{Ls}
\end{align}
In the following we will focus on the quantum problem at zero temperature, when Eqs. (\ref{Ls}) become:

\begin{align} &L_1(t)=\pi g^2 \sin \omega_0 t+4\pi \alpha \omega_c^3 \frac{\omega_c t}{\left(1+\omega_c^2t^2\right)^2} \notag \\
&L_2(t)=\pi g^2 \cos \omega_0 t+2\pi \alpha \omega_c^2 \frac{1-\omega_c^2 t^2}{\left(1+\omega_c^2t^2\right)^2}.
\end{align}

\begin{figure}
\center
\includegraphics[scale=0.30]{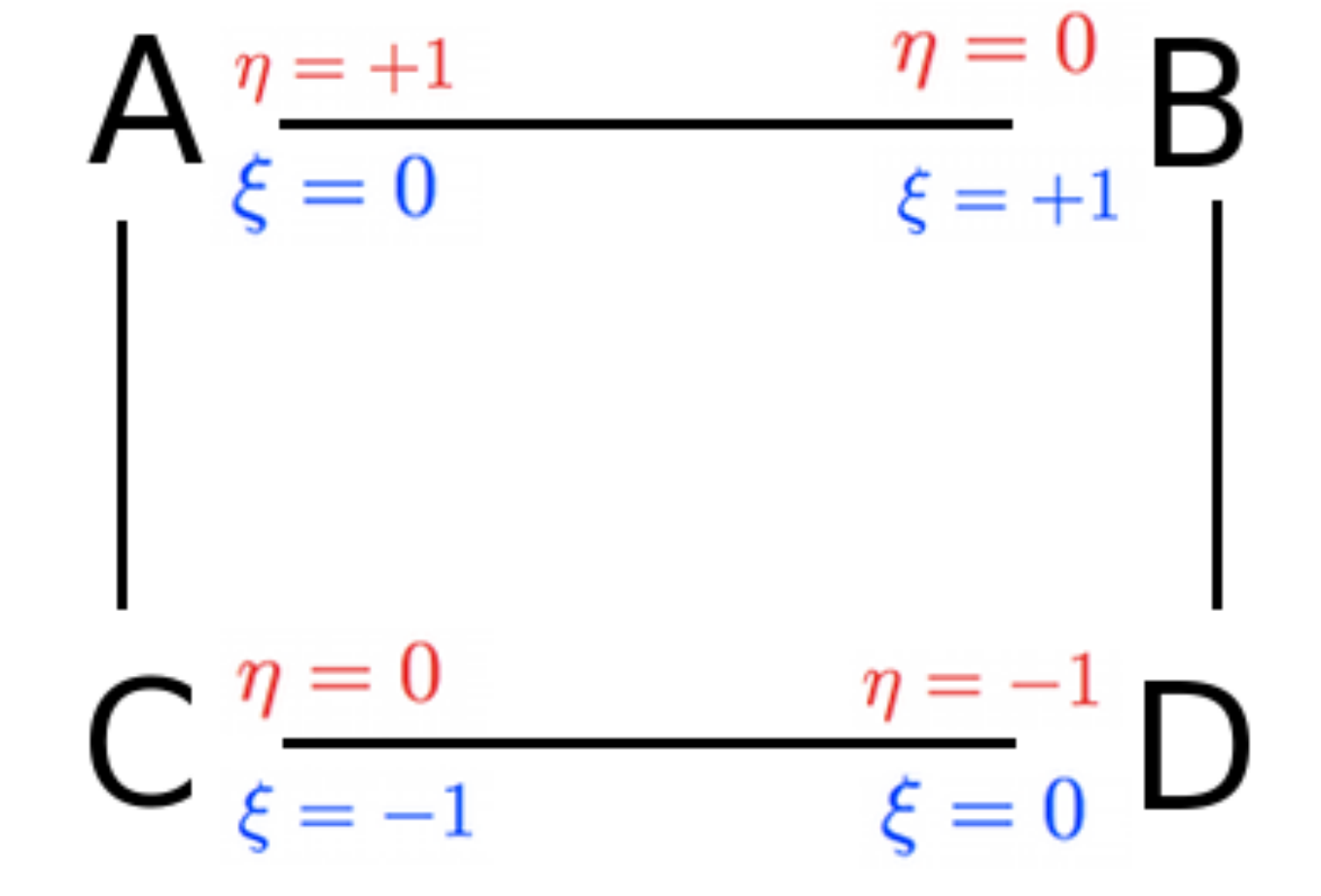}  
\caption{(Color online) States of the double spin path. Through the Feynman-Vernon formulation, the spin paths and elements of the spin (reduced) density matrix are depicted through the classical variables $\eta$ and $\xi$.}
\label{etats}
\end{figure}

 The double path integral along $\sigma$ and $\sigma'$ in Eq. (\ref{eq:densitymatrixelement}) can be viewed as a single path that visits the four states $\textrm{A}=|++\rangle$, $\textrm{B}=|+- \rangle$, $\textrm{C}=|-+ \rangle$ and $\textrm{D}=|--\rangle$ (see Fig. 1). States A and D correspond to the diagonal elements of the density matrix (also named `sojourn' states) whereas B and C correspond to the off-diagonal ones (also called `blip' states) \cite{Leggett,weiss,Lesovik}. More precisely, we have to take into account all possible paths along the edges of the square in Fig. \ref{etats}. In order to go further, we have to fix the initial and final states of such paths.
 
We will first focus on the computation of the diagonal elements of the density matrix, $(1+\langle\sigma^x \rangle)/2$, when the spin is initially in a pure state along the $x$-axis $\rho_S(t_{0})=|+_x\rangle \langle +_x|$. The computation of off-diagonal elements of the density matrix will be addressed in susbsection B, and the preparation $\rho_S(t_{0})=|+_z\rangle \langle +_z |$ in subsection C.

\subsection{Diagonal elements of the spin density matrix}

The initial condition is $\rho_S (t_0)=|+_x\rangle \langle +_x| $, so that the double spin path is initially constrained in the diagonal state $\textrm{A}=|++\rangle$. We intend to compute the upper left diagonal element of the density matrix describing the probability to find the system in the state $|+_x\rangle $ at time $t$, so that we consider spin paths that end in the sojourn state $\textrm{A}=|++\rangle$. One path of this type makes $2n$ transitions along the way at times $t_i$, $i \in \{1,2,..,2n\}$ such that $t_0<t_1<t_2<...<t_{2n}$. We can write this spin path as $\xi(t)=\sum_{j=1}^{2n} \Xi_j\theta(t-t_j)$ and $\eta(t)=\sum_{j=0}^{2n} \Upsilon_j\theta(t-t_j)$ where the variables $\Xi_i$ and $\Upsilon_i$ take values in $\{-1,1\}$. Such a path is illustrated in Fig. \ref{spin_path_1}. The variables $\Xi$ (in blue) describe the blip parts, and the variables $\Upsilon$ (in red) on the other hand characterize the sojourn parts. 

The diagonal element of the density matrix 
\begin{equation}
p(t)=\frac{1+\langle \sigma^x(t)\rangle}{2} =\frac{1+{\cal P}(t)}{2},
\end{equation}
is given by a series in the tunneling coupling $\Delta^2$ \cite{Leggett,weiss,stochastic} :
\begin{equation}
p(t)=\sum_{n=0}^{\infty} \left(\frac{i\Delta}{2} \right)^{2n} \int_{t_0}^{t} dt_{2n} ... \int_{t_0}^{t_2} dt_{1} \sum_{\{\Xi_j\},\{\Upsilon_j\}' } \mathcal{F}_{n}.
\label{eq:p(t)}
\end{equation}
 The prime in $\{\Upsilon_j\}'$ in Eq. (\ref{eq:p(t)}) indicates that the initial and final sojourn states are fixed according to the initial and final conditions, $\Upsilon_0=\Upsilon_{2n}=1$. Therefore: 
 
  \begin{figure}[t!]
\center
\includegraphics[scale=0.42]{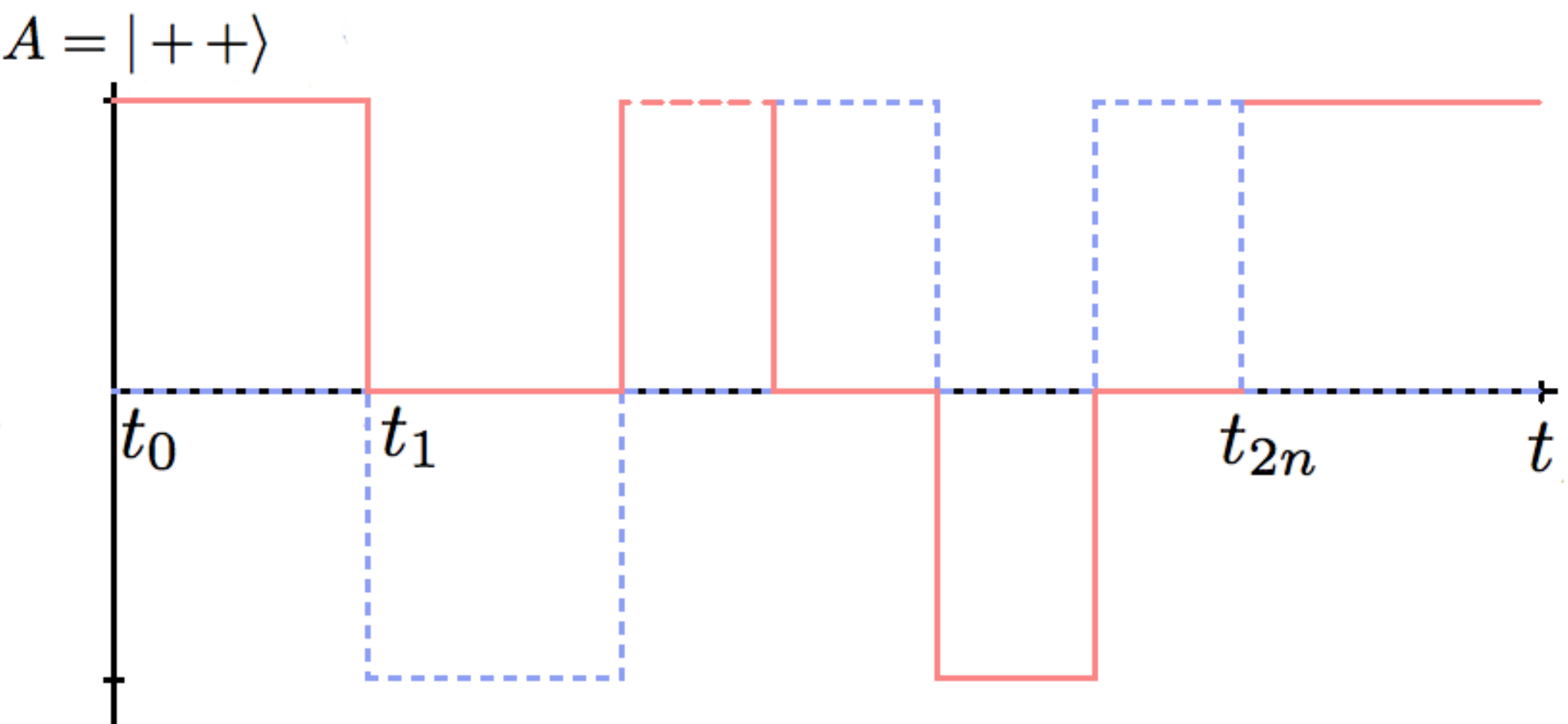}  
\caption{(Color online) Spin path $\eta(t)=\sum_{j=0}^{2n} \Upsilon_j\theta(t-t_j)$ in red; $\xi(t)=\sum_{j=1}^{2n} \Xi_j\theta(t-t_j)$ in dashed blue. The spin path starts and ends in the state A.}
\label{spin_path_1}
\end{figure}
 
 \begin{align}
 &\mathcal{F}_{n}= \mathcal{F}_{n}[\{\Xi_j\},\{\Upsilon_j\}, \{ t_j\}] = \mathcal{Q}_1 \mathcal{Q}_2 \label{10}\\
  &\mathcal{Q}_1 =\exp \left[ \frac{i}{\pi} \sum_{k=0}^{2n-1}\sum_{j=k+1}^{2n} \Xi_j \Upsilon_k  Q_1(t_j-t_k) \right] \label{Q_1} \\
  &\mathcal{Q}_2 =\exp \left[ \frac{1}{\pi} \sum_{k=1}^{2n-1}\sum_{j=k+1}^{2n} \Xi_j \Xi_k  Q_2(t_j-t_k) \right]\label{Q_2} . 
 \end{align}
 
The functions $Q_1$ and $Q_2$, which describe the feedbacks of the electromagnetic field and of the dissipative environment, are directly obtained from the spectral function $J(\omega)$. At zero temperature, they read:
 \begin{align}
 Q_1(t)&=\pi \left[ \frac{g^2}{\omega_0^2} \sin \omega_0 t + 2 \alpha \tan^{-1} (\omega_c t) \right] \notag \\
 Q_2(t)&=\pi \left[ \frac{g^2}{\omega_0^2} (1-\cos \omega_0 t) +  \alpha \ln (1+\omega_c^2 t^2) \right].
 \label{Q_expression}
\end{align}
The $\ln$-function in $Q_2$ reflects the non-Markovian features of the ohmic bath \cite{Stockburger}.
 It is important to notice that blips and sojourns do not have symmetric effects. $\mathcal{Q}_1$ describes the coupling between the blips and all previous sojourns, and $\mathcal{Q}_2$ countains the interaction between all blips (including self interaction). The index for the $\Upsilon$ variables starts at $0$ and ends at $2n-1$ whereas the index for the $\Xi$ variables starts at $1$ and ends at $2n$. It is worth  noting that the last sojourn does not contribute and the latest coupling period is the blip which lasts from $t_{2n-1}$ to $t_{2n}$. 
 
Let now $h_{\xi}$ and $h_{\eta}$ be two complex gaussian random fields which verify:
\begin{align}
 \overline{ h_{\xi}(t) h_{\xi}(s)} = & \frac{1}{\pi} Q_2(t-s) + k_1 \notag\\
 \overline{ h_{\eta}(t) h_{\eta}(s)} = &\  k_2      \notag\\
 \overline{ h_{\xi}(t) h_{\eta}(s) } = & \frac{i}{\pi}  Q_1(t-s) \theta(t-s) + k_3.
 \label{heightfunctions}
\end{align}
The overline denotes statistical average, and $k_1$, $k_2$ and $k_3$ are arbitrary complex constants. Making use of the identity $\overline{\exp(X)}=\exp(\overline{X^2}/2 )$ , Eqs. (\ref{10}), (\ref{Q_1}) and (\ref{Q_2}) can then be reexpressed as:

 \begin{align}
\mathcal{F}_{n}= \overline{  \prod_{j=1}^{2n} \exp\left[ h_{\xi}(t_j) \Xi_j+h_{\eta}(t_{j-1}) \Upsilon_{j-1}    \right]}.
\label{functionnal_1}
\end{align}
The complex constants $k_p$ do not contribute because $\sum_{k=0}^{2n-1} \Upsilon_k=\sum_{j=1}^{2n} \Xi_j=0$. Practically fields which verify correlation relations (\ref{heightfunctions}) can be sampled by Fourier series decomposition (see Appendix \ref{Appendix_A} for more details). 

From Eqs. (\ref{eq:p(t)}) and (\ref{functionnal_1}), the resulting formula for $p(t)$ can be expressed as:
\begin{equation} 
p(t) = \sum_{n=0}^{\infty} (-1)^n  \int_{t_0}^{t} dt_{2n}..\int_{t_0}^{t_2} dt_{1} \langle \Phi_f | W(t_{2n})...W(t_1)  |\Phi_i \rangle,
\label{final2}
\end{equation}
where the effective Hamiltonian $W(t)$ for the spin in the four-dimensional vector space of states $\{|++ \rangle , |+-\rangle, |-+\rangle, |--\rangle \}$ is:

\begin{equation}
W(t)=\frac{\Delta }{2} \left( \begin{array}{cccc}
0&e^{-h_{\xi}+h_{\eta} }&-e^{h_{\xi}+h_{\eta} }&0 \\
e^{h_{\xi}-h_{\eta} }&0&0&-e^{h_{\xi}+h_{\eta} }\\
-e^{-h_{\xi}-h_{\eta} }&0&0&e^{-h_{\xi}+h_{\eta} }\\
0&-e^{-h_{\xi}-h_{\eta} }&e^{h_{\xi}-h_{\eta} }&0
\end{array} \right).
\label{eq:system}
\end{equation}
We have $|\Phi_i \rangle=(e^{h_{\eta}(t_0)},0,0,0)^T$  and $\langle \Phi_f |=(e^{-h_{\eta}(t_{2n})},0,0,0)$: these choices account for the asymmetry between blips and sojourns. The contribution from the first sojourn is encoded in $|\Phi_i \rangle$, and we artificially suppress the contribution of the last sojourn via $|\Phi_f \rangle$. This final vector depends on the intermediate time $t_{2n}$, but we can notice that replacing $(e^{-h_{\eta}(t_{2n})},0,0,0)$ by $(e^{-h_{\eta}(t)},0,0,0)$ does not add any contribution on average. Then we can write $p(t)$ as a time-ordered product:
\begin{equation} 
p(t) = \overline{\langle \Phi_f | \mathcal{T} e^{-i\int_{t_0}^{t}ds W(s)} |\Phi_i \rangle},
\end{equation}
where $\mathcal{T}$ is the time-ordering operator. Resorting to Eq. (2), we rewrite $p(t)$ as a stochastic average $\overline{ \langle \Phi_f |  \Phi(t) \rangle}$, where $| \Phi(t) \rangle$ is the solution of the Sch\"{o}dinger equation
\begin{equation} 
i\partial_{t}|\Psi(t) \rangle=W(t)|\Psi(t) \rangle 
\label{SSE}
\end{equation}
with the initial condition $|\Phi_i \rangle$.

It is actually possible to consider photon leakage out of the cavity when $\Gamma \neq 0$. This change can be treated exactly in the integration and just leads to a change in the coupling functions  $Q_1$ and $Q_2$. The first terms of the right hand side of Eq. (\ref{Q_expression}) are multiplied by a damping factor $\exp\left(-\Gamma t\right)$.


\subsection{Off-diagonal elements of the spin density matrix}  
  
\begin{figure}[t!]
\center
\includegraphics[scale=0.42]{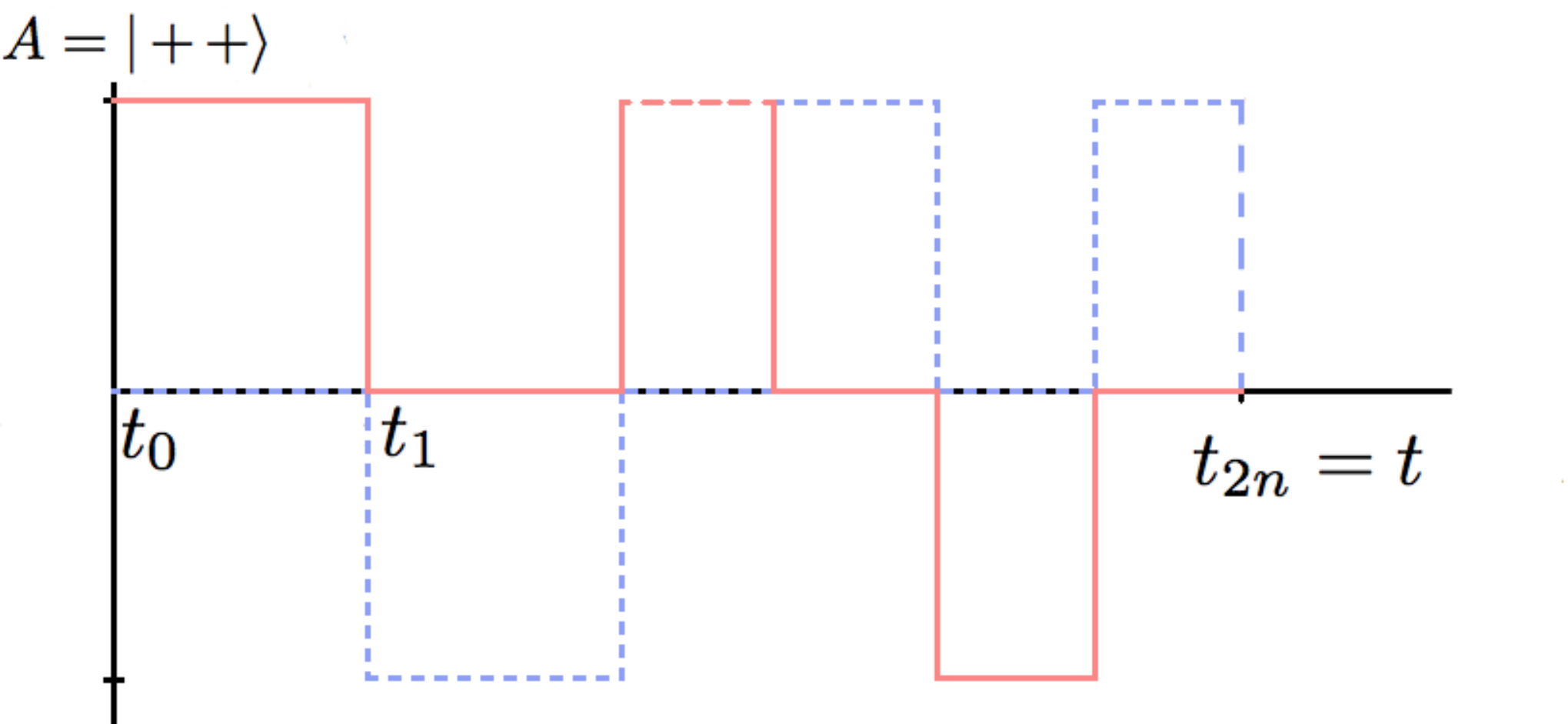}  
\caption{(Color online) Spin path- $\eta(t)=\sum_{j=0}^{2n-1} \Upsilon_j\theta(t-t_j)$ in red; $\xi(t)=\sum_{j=1}^{2n-1} \Xi_j\theta(t-t_j)$ in dashed blue. Here the spin path ends in the blip state $\textrm{B}=|+- \rangle$. The initial state is still $\textrm{A}=|++ \rangle$.}
\label{spin_path_2}
\end{figure}

Following the work by Weiss \cite{weiss}, we compute an off-diagonal term of the density matrix in terms of a series expansion in $\Delta$, considering spin paths that end in a blip state. Such paths make now $2n-1$ transitions and we have:
\begin{equation}
\rho_{+-} (t)=\sum_{n=1}^{\infty} \frac{(i\Delta)^{\small{2n-1}}}{2^{\small{2n-1}}} \int_{t_0}^{t} dt_{\small{2n-1}} ... \int_{t_0}^{t_2} dt_{1}\sum_{\{\Xi_j\}'\{\Upsilon_j\}' }   \mathcal{F}_{n},
\label{eq:off(t)}
\end{equation}
where $\rho_{+-} (t)=\langle +|\rho_S (t)|-\rangle$. An example of such a path can be seen in Fig. \ref{spin_path_2}. Here the initial sojourn state is fixed, as well as the final blip state.
All blips are coupled to all previous sojourns and blips as can be seen in Eqs. (\ref{Q_1}) and (\ref{Q_2}). Paths considered in subsection A ended in a sojourn state, and the latest coupling period lasted from $t_{2n-1}$ to $t_{2n}$. The situation is different here because paths end up in a blip state. For a given path the final coupling period then lasts from $t_{2n-1}$ to the final time $t$. But providing that we formally set $t_{2n}=t$ and $\Xi_{2n}=-\sum_{j=1}^{2n-1}\Xi_{j}$, equations (\ref{10}), (\ref{Q_1}) and (\ref{Q_2}) are still valid. Following the same route, we can conclude that $\langle \sigma^z (t) \rangle$ is given by the average $ 2 \textrm{Re} \overline{ \langle \Phi'_f | \Phi'(t) \rangle}$ where $| \Phi' (t)\rangle$ is the solution of the stochastic Schr\"{o}dinger equation with the initial condition $|\Phi'_i \rangle=|\Phi_i \rangle$ and  $\langle\Phi'_f |=(0,e^{-h_{\xi}(t)},0,0)$.

 \subsection{Initial condition}  
 
 It is also possible to consider a protocol in which the spin is initially prepared in an eigenstate along the $z$-axis:  $\rho_S(t_{0})=|+_z\rangle \langle +_z|$. Due to the linearity of Eq. (\ref{eq:densitymatrixelement}) we can evolve the four initial components of the density matrix separately.  The treatment of the evolution of a diagonal element of the density matrix  have already been done in subsection A (the case of a path begining in the $\textrm{D}=|--\rangle$ state can be deduced considering $\Upsilon_0=-1$). We focus then on the evolution of a path which is initially in a blip state. We have:
\begin{equation}
\rho_{+-} (t)=\sum_{n=0}^{\infty} \left(\frac{i\Delta}{2} \right)^{2n} \int_{t_0}^{t} dt_{2n} ... \int_{t_0}^{t_2} dt_{1}\sum_{\{\Xi_j\}'} \sum_{\{\Upsilon_j\} }   \mathcal{F}_{n}.
\end{equation} 
\begin{figure}[t!]
\center
\includegraphics[scale=0.42]{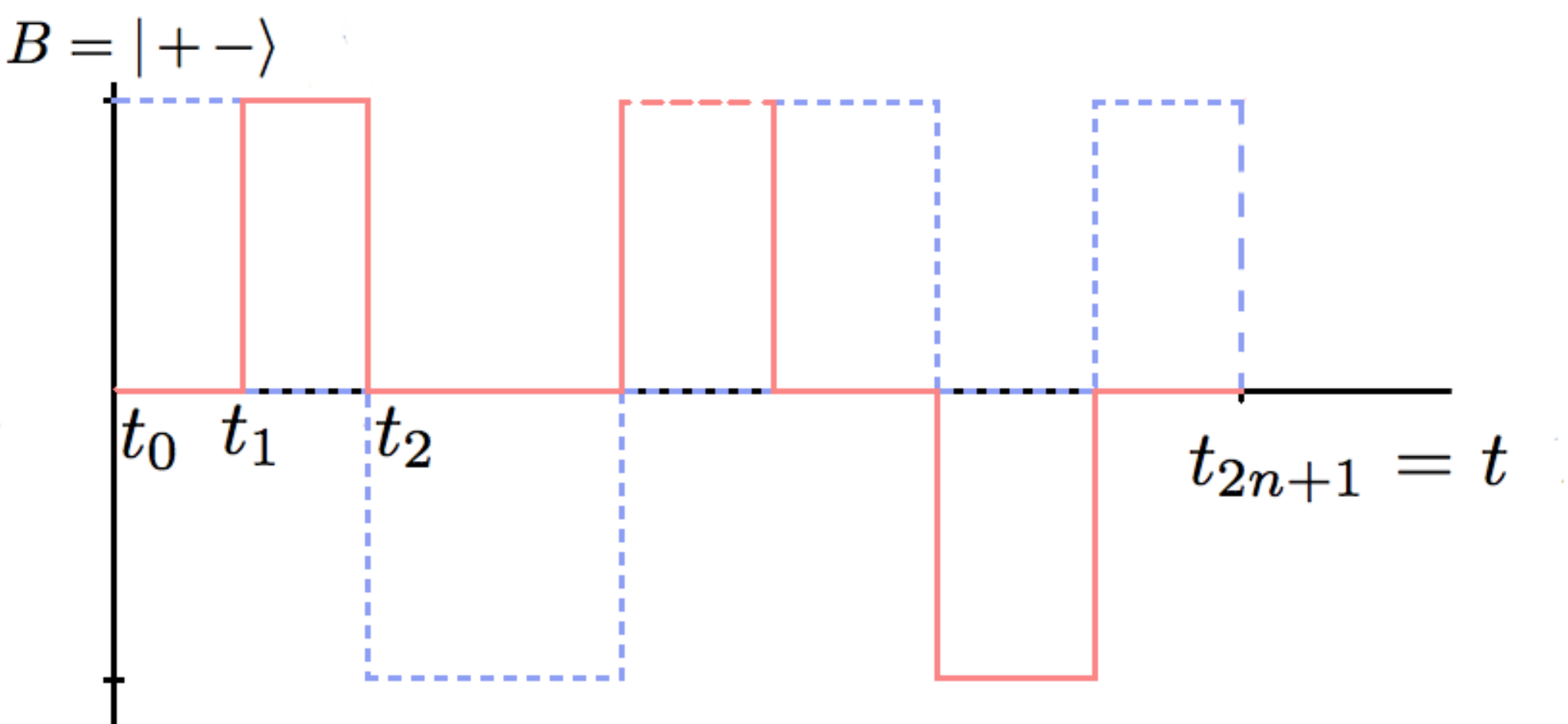}  
\caption{(Color online) Spin path- $\eta(t)=\sum_{j=1}^{2n} \Upsilon_j\theta(t-t_j)$ in red; $\xi(t)=\sum_{j=0}^{2n} \Xi_j\theta(t-t_j)$ in dashed blue. Here the spin path starts and ends in the blip state $\textrm{B}=|+-\rangle$.}
\label{spin_path_3}
\end{figure}  
Here initial and final blip states are constrained, as shown in Fig. \ref{spin_path_3}. We formally set $t_{2n+1}=t$ and considering for example a path that starts in state B, we find:  

\begin{align} \mathcal{F}_{n}=&\exp\left[ \frac{i}{\pi} \sum_{k=1}^{2n}\sum_{j=k+1}^{2n+1} \Xi_j \Upsilon_k  Q_1(t_j-t_k)\right]\notag \\
 &\times \exp \left[ \frac{1}{\pi} \sum_{k=0}^{2n}\sum_{j=k+1}^{2n+1} \Xi_j \Xi_k  Q_2(t_j-t_k) \right]. \label{Fbis}
  \end{align}
 
Then the expression of $\langle \sigma^z (t) \rangle$ is given by:
\begin{equation} 
\langle \sigma^z (t) \rangle= 2 \textrm{Re} \sum_{n=0}^{\infty}  \int_{t_0}^{t} dt_{2n} ... \int_{t_0}^{t_2} dt_{1}  \chi (t_1, t_2,...,t_{2n}) ,
\end{equation}
where  $\chi(t_1, t_2,...,t_{2n})=\overline{ \langle \Phi''_f | W(t_{2n})...W(t_2)W(t_1)  |\Phi''_i \rangle      }$, $|\Phi''_i \rangle=(0,e^{h_{\xi}(t_0)},0,0)^T$, $\langle\Phi''_f |=\langle\Phi'_f |$. 

\subsection{External drive}
 The effect of a coherent semi-classical drive of the form $V(t) (a+a^{\dagger})$ can be treated exactly by substituing $\sigma^x (t)$ by $(\sigma^x (t)+V(t))$ in the path integral approach. This is simply reflected in Eq. (\ref{eq:influence}) by the appearance of a new coupling term. Assuming $V(t)$ to be of the form $V_0 \cos \omega_d t$ and beginning the procedure at time $t_0$, the functional $F_{\sigma, \sigma'}$ is changed into $F^d_{\sigma, \sigma'}$ which reads, for $t \geq t_0$:

\begin{equation}
 F^d_{\sigma, \sigma'}=e^{\left[ 2 i V_0 g \int_{t_0}^t ds \int_{t_0}^s ds' \sin \omega_0 (s-s') \xi(s)  \cos \omega_d s'    \right]} F_{\sigma, \sigma'}. 
 \end{equation}
We consider for example a path that starts in a sojourn state and ends in a blip state. The new contribution can be taken into account into one height field. Let us call $h_{\xi}^d$ the stochastic field coupling blips in the presence of the drive. It reads:
\begin{align}
\label{hxi_drived}
h_{\xi}^d(t)=h_{\xi}(t)+\frac{2iV_0 g \omega_0}{\omega_d^2-\omega_0^2} \Bigg\{&\frac{\sin \left[\omega_0 t+(\omega_0+\omega_d) t_0\right]}{\omega_0} \notag \\
&+\frac{ \sin \omega_d t}{\omega_d}\Bigg\}.
\end{align}
It is also possible to consider the drive term with a RWA-type approximation $V_0/2 \left(ae^{i\omega_d t}+a^{\dagger}e^{-i\omega_d t}\right)$, which only results in the replacement of $2 V_0 g  \omega_0/(\omega_d^2-\omega_0^2)$ by $V_0 g/(\omega_d-\omega_0)$ in Eq. (\ref{hxi_drived}). 

In this subsection we have considered the effect of an external photonic drive. It is also possible to incorporate the effect of drive term acting on the qubit $\epsilon(t) \sigma^x$, as in Ref. \cite{2010stoch}. The stochastic field then reads:
\begin{align}
h_{\xi}^d(t)=h_{\xi}(t)+\int_{t_0}^t ds~ \epsilon (s).
\end{align}
This makes connections with the Landau-Zener-Majorana-St\"uckelberg oscillations \cite{Nori,petta}, which can studied with this approach.

In this section we have shown that it is possible to compute the dissipative and driven dynamics of the spin in an exact manner by evaluating a stochastic Schr\"{o}dinger for the spin-reduced density matrix, and the effect of the environment is fully encapsulated in the correlations of the random Hamiltonian of the stochastic equation.

\section{Dynamics of the Rabi Model}

In this Section, we will focus on the spin dynamics without external drive, and study the free Rabi model and the deviations from the Jaynes-Cummings dynamics.

\subsection{Corrections to the Jaynes-Cummings model}

\begin{figure}[t!]
\center
\includegraphics[scale=0.4]{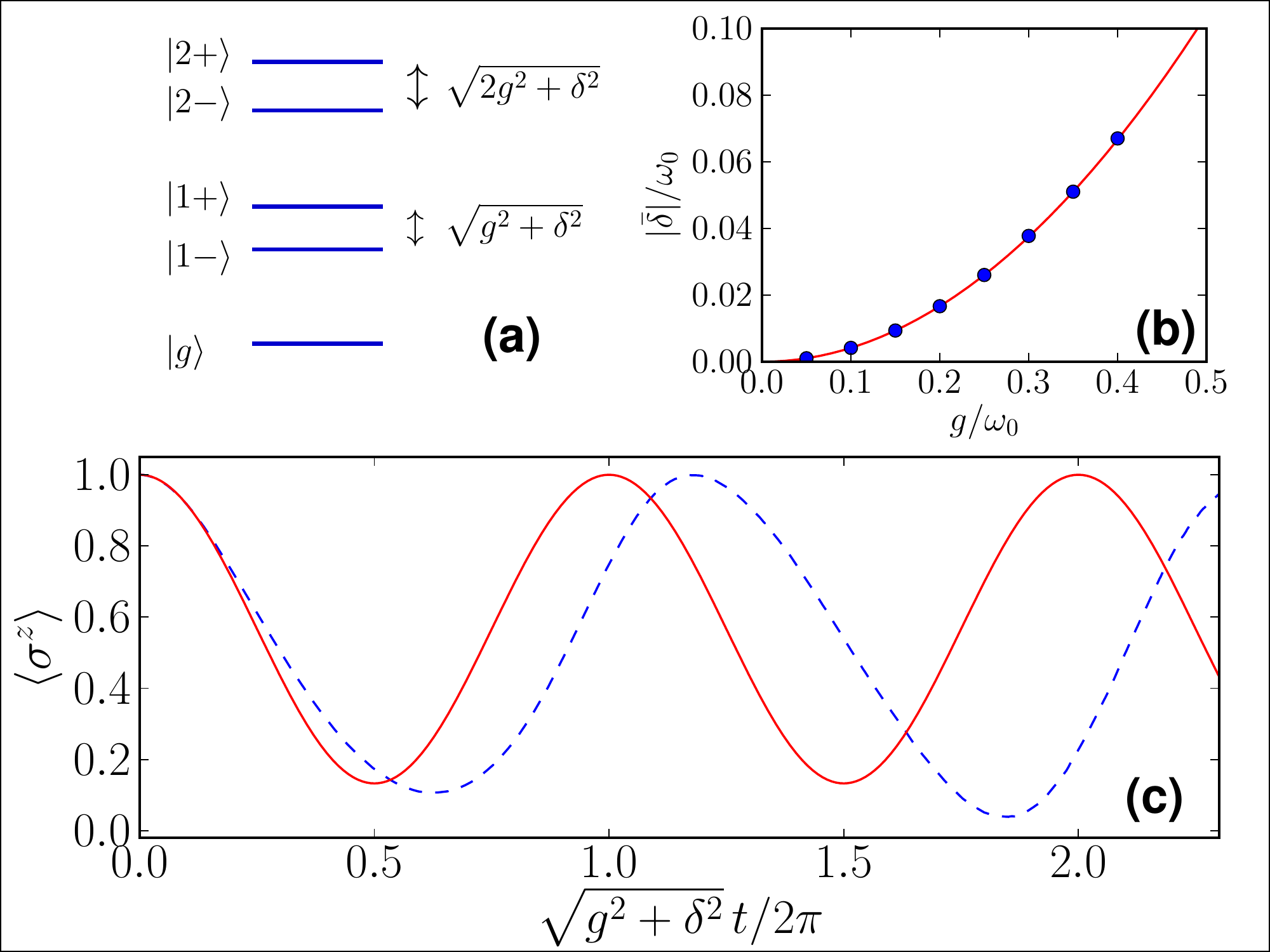}  
\caption{(Color online) (a) JC ladder of polaritons. (b) Absolute value of the Bloch-Siegert shift $|\bar{\delta}|$ versus $g/\omega_0$, perturbation theory in red \cite{cohen_2}, and results from our method in blue (dots). (c) Example of dynamics of $\langle \sigma^z \rangle$ with the initial condition $|+_z\rangle$, which is a linear superposition of $|1-\rangle$ and $|1+\rangle$, for quite strong couplings. Parameters are set to $g/\omega_0=0.7$, $\Delta/\omega_0=0.2$. The Rabi solution from our method is shown in dashed blue; within the RWA, the JC solution would rather read $\langle\sigma^z(t)\rangle=1-2g^2\sin^2(\sqrt{g^2+\delta^2}\, t/2)/(g^2+\delta^2)$ and is shown in red.}
\label{bloch_siegert}
\end{figure}

First, we check that the results for the free Rabi model reproduce the dynamics of the JC model in the weak coupling limit $g/\omega_0 \ll 1$ with weak detuning $\delta/\omega_0\ll 1$ where $\delta=\omega_0-\Delta$. Since the RWA holds we can easily diagonalize the free undamped JC Hamiltonian in the so-called dressed basis. The ground state $|g \rangle$ of the system consists of the two-level system in its lower state and vacuum for the photons, while the excited eigenstates $|n{\pm}\rangle$ are pairs of combined light-matter excitations (polaritons) described in terms of the polariton number operator $N=a^{\dagger} a + \sigma^+\sigma^-$ which commutes with the Hamiltonian, $N|n_{\pm}\rangle=n|n_{\pm}\rangle$. This leads to the well-known structure of the anharmonic JC ladder (Fig. \ref{bloch_siegert}(a)). More precisely, the light-matter eigenstates satisfy (here, $n>0$): 
\begin{align}
 &|g \rangle= |-_z,0\rangle  \\ \notag 
 &|n+\rangle=  \alpha_n  |+_z,n-1\rangle+\beta_n  |-_z,n\rangle \\ \notag         
 &|n-\rangle=     -\beta_n  |+_z,n-1\rangle+\alpha_n  |-_z,n\rangle.
\end{align}
The corresponding energies are:
\begin{align}
 &E_{|g \rangle}=\frac{\delta}{2}  \\ \notag 
 &E_{|n+\rangle}=n \omega_0   +\frac{1}{2}\sqrt{\delta^2+ n g^2}      \\ \notag         
 &E_{|n-\rangle}= n \omega_0 -\frac{1}{2}\sqrt{\delta^2+ n g^2}.
 \label{JC_ref}
\end{align}
We have: $\alpha_n=\sqrt{\left[A(n)-\delta\right]/2A(n)}$, $\beta_n=\sqrt{\left[A(n)+\delta\right]/2 A(n)}$, and $A(n)=\sqrt{ng^2+\delta^2}$. 

If one prepares the two-level system in its upper state $|+_z\rangle$ in vacuum, the dynamics shows coherent oscillations between the two polariton states $|1-\rangle$ and $|1+\rangle$, also known as Rabi oscillations. We obtain consistent results, and the first corrections to the JC limit when the coupling $g$ is increased, are in agreement with (second order) perturbation theory (Fig. \ref{bloch_siegert}(b)). The presence of the counter-rotating terms in the quantum Rabi model gives rise to a shift of the resonance frequency between the atom and photon, leading to an additional negative detuning $\bar{\delta} = -g^2/\left[2(\omega_0+\Delta)\right]$ when $\Delta<\omega_0$. This Bloch-Siegert shift \cite{cohen_2} has been observed in circuit QED \cite{Mooij}. Moreover, the dynamics towards the deep strong coupling regime with $g\approx\omega_0$ certainly goes beyond this perturbative argument \cite{braak2,solano}, as shown in Fig. \ref{bloch_siegert}(c).

\begin{figure}[t!]
\center
\includegraphics[scale=0.4]{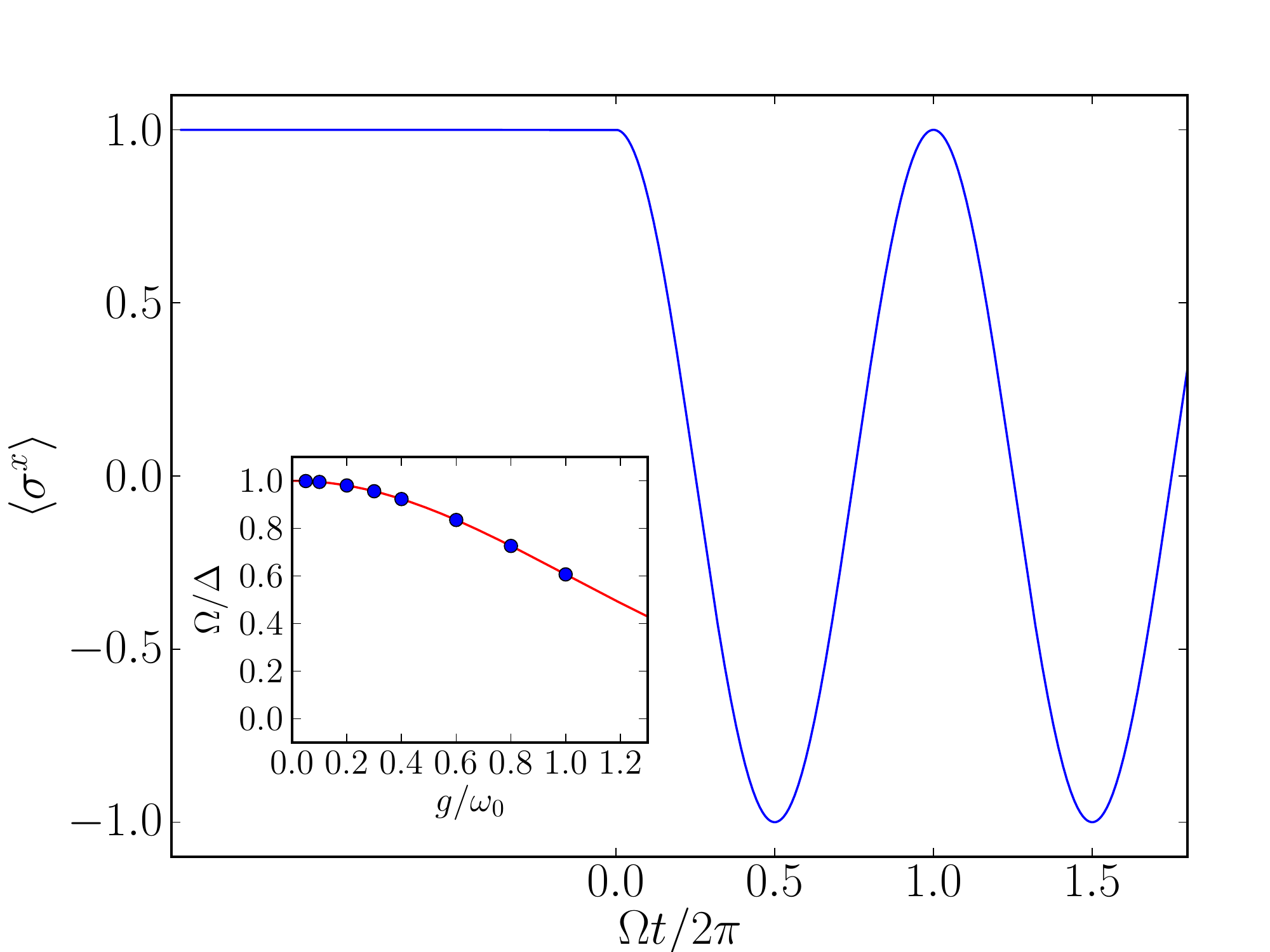}  
\caption{(Color online) Dynamics of $\langle \sigma^x \rangle$ with the initial condition $|+_x\rangle$, for $\Delta/\omega_0=0.1$ and  $g/\omega_0=0.3$. The analytical solution within the adiabatic approximation is $\langle \sigma^x(t) \rangle = \cos \Omega t$. Inset: $\Omega/\Delta=\exp(-g^2/2\omega_0^2)$ versus $g/\omega_0$; the red curve is from the exact result in the adiabatic limit while the dots represent our numerical results.}
\label{adiabatic}
\end{figure}

\subsection{Adiabatic limit}
The regime of the Rabi model  corresponding to a highly detuned system with $\Delta/\omega_0 \ll 1$ is known as the adiabatic limit \cite{adiabatic_0}. One can visualize such a system as a set of two displaced oscillator wells (characterized by the value of $\sigma^x$), whose degeneracy is lifted by the field along the $z$-direction. The dynamics of the two-level system, initially prepared in a displaced state of one well, should undergo coherent and complete oscillations between this state and its symmetric counterpart in the other well. Such an initial state can be prepared by applying a strong bias field along $x$-direction for negative times, letting the system relax towards its shifted equilibrium position before the release of the constraint at time $t=0$. The frequency of oscillations only depends on the overlap between these two states, and one can show that this frequency is $\Omega=\Delta e^{-g^2/2 \omega_0^2} $ \cite{adiabatic_1} (Fig. \ref{adiabatic}). 

\begin{figure}[h!]
\center
\includegraphics[scale=0.4]{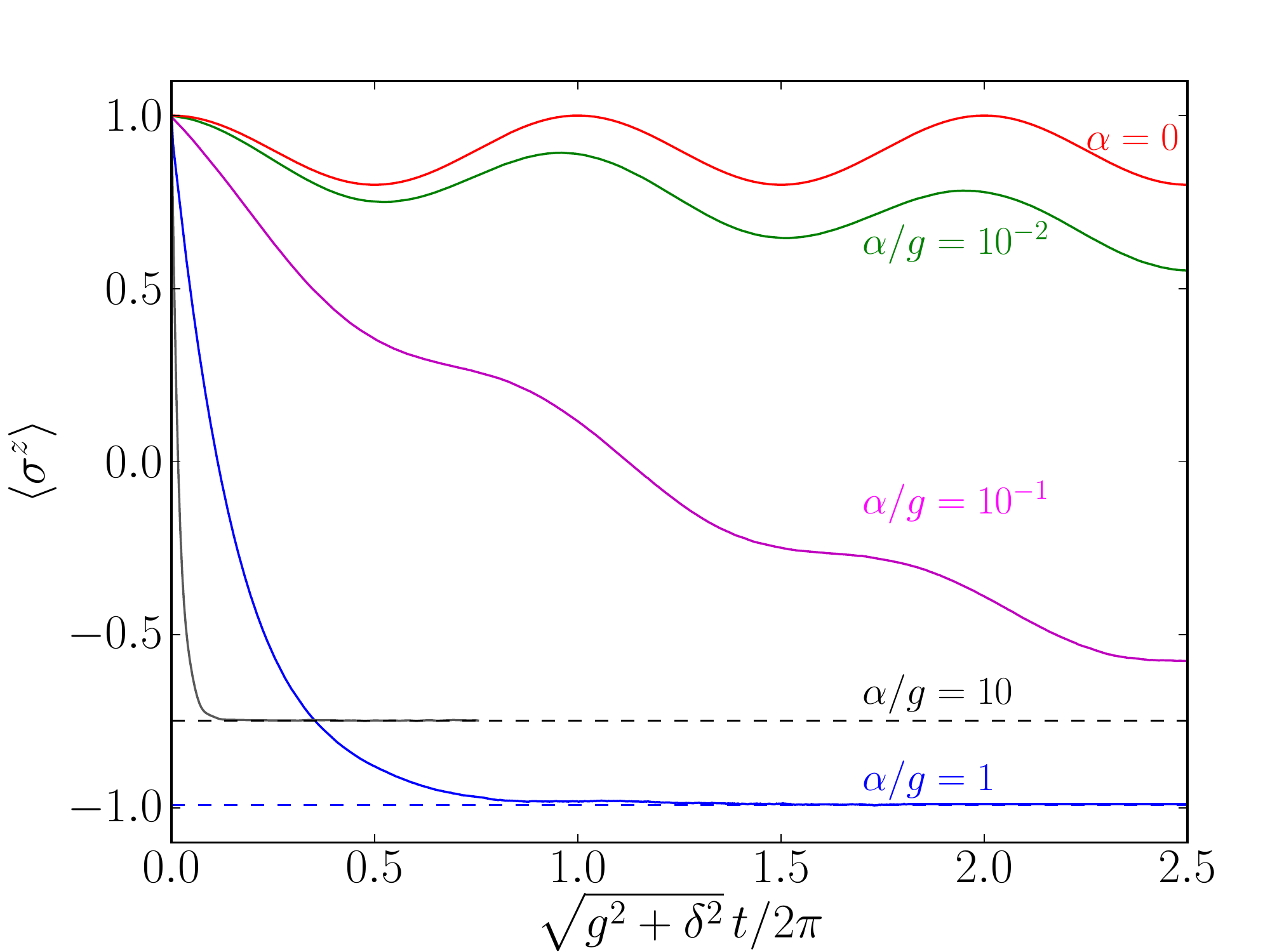}  
\caption{(Color online) Dynamics of $\langle \sigma^z \rangle$ with the initial condition $|+_z\rangle$ which is a linear superposition of the first two polaritons. We consider $g/\omega_0=0.01$, $\Delta/ \omega_0=0.97$, $\omega_c/\omega_0=100$ and several values of $\alpha$ until $\alpha\approx 0.1$. We observe a relaxation towards a non-trivial final state by increasing $\alpha$ and the value of $\langle \sigma^z \rangle$ is in accordance with Bethe Ansatz calculations  \cite{Filyov,Cedraschi1,buttiker,kopp,KLH}.}
\label{damped_1}
\end{figure}

The convergence of the numerical evaluation in our procedure is ensured from weak coupling to ratios $g/\omega_0$ of the order of $1$, allowing us to reach the ultra-strong coupling regime. 

\section{Dissipation, drive and lattice}
In this Section, we go one step further and consider dissipation and drive effects non-perturbatively. Lattice effects will then be briefly addressed.

\subsection{Dissipation effects} 

Decoherence effects are characterized by a prominent suppression of the off-diagonal elements of the spin reduced density matrix at equilibrium \cite{buttiker,kopp,KLH} as well as a damping (disappearance) of the Rabi oscillations \cite{Leggett,weiss,stochastic,anders,bulla,david_peter,kennes,schoeller,Lesage,loss} (see Fig. \ref{damped_1}). More precisely, the effects of the ohmic bath at low coupling are both a damping of the Rabi oscillations and a dephasing due to a renormalization of the tunneling element  $\Delta$ to $\Delta_r=\Delta (\Delta/\omega_c)^{\alpha/1-\alpha}$ \cite{Leggett,weiss}. The latter phenomenon engenders an effective detuning $\delta_r=\omega_0-\Delta_r$ between the two-level system and light, which can be seen explicitly via a change in the frequency of Rabi oscillations in the dynamics of $\langle \sigma^z \rangle$ and $\langle \sigma^x \rangle$. The numerical estimation of this effective detuning matches the theoretical expectation $\delta_r$ for coupling strengths $\alpha/g \ll 1$. 

\begin{figure}[h!]
\center
\includegraphics[scale=0.4]{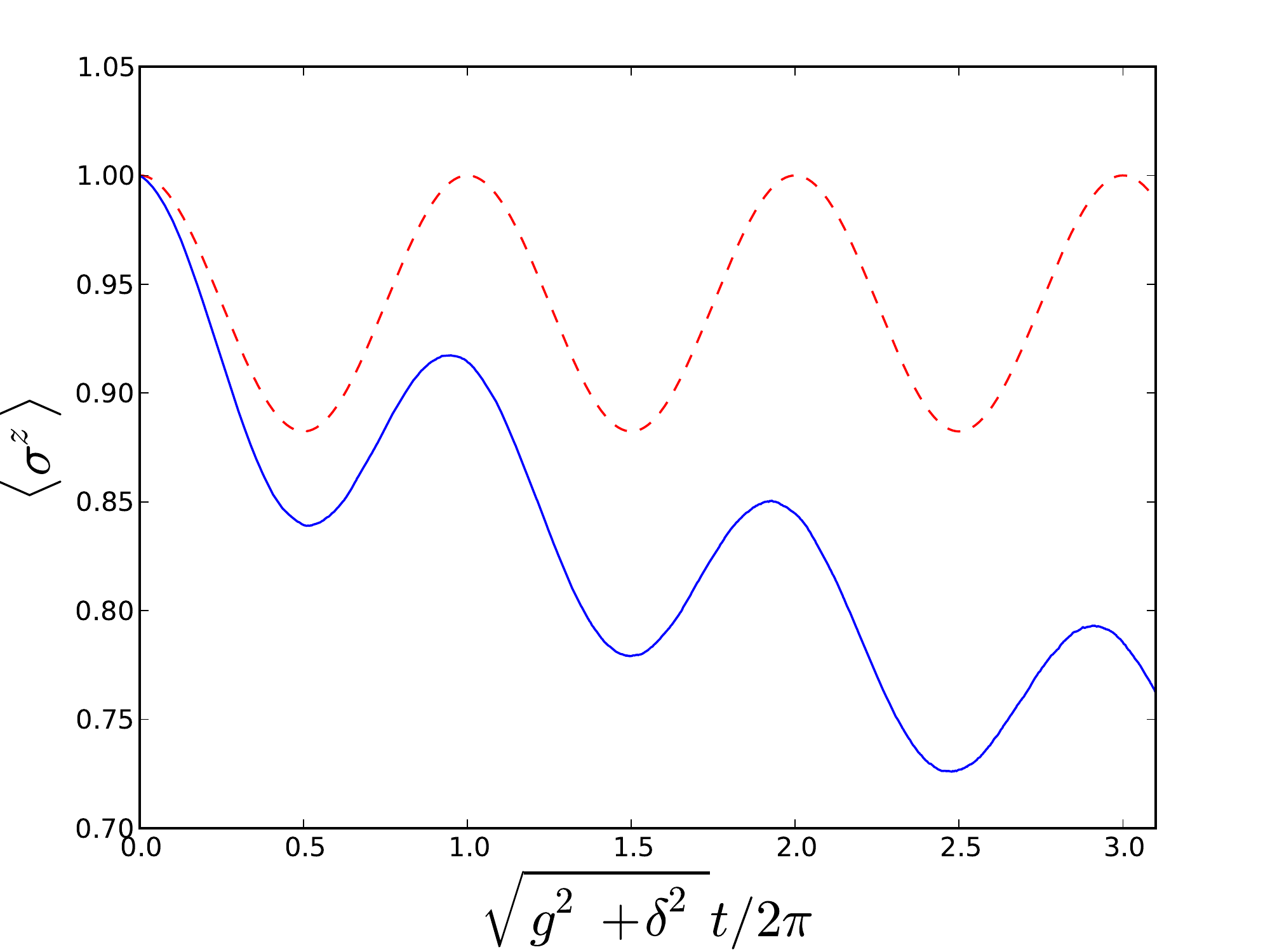}  
\caption{(Color online) Dynamics of $\langle \sigma^z \rangle$ with the initial condition $|+_z\rangle$. The blue curve shows the damped evolution in presence of photon leakage out of the cavity whereas the dashed red curve represents the undamped case. Parameters are $g/\omega_0=5.0~10^{-2}$, $\Delta/\omega_0=0.8$, $\Gamma/\omega_0=1.0~10^{-5}$ and $\alpha=0$.}
\end{figure}   

Note that $\Delta_r$ can also be identified as the effective Kondo energy scale in the ohmic spin-boson model \cite{Leggett,weiss}. When the dissipation strength increases the net field along the $x$-axis progressively becomes zero, since $\langle a+a^{\dagger}\rangle \approx 0$ at $g/\alpha \ll1$. Then, the system relaxes to a final state with $\langle \sigma^x\rangle=0$, and $\langle \sigma^z\rangle$ (within our notations) can be evaluated through Bethe Ansatz calculations \cite{Filyov,Cedraschi1,buttiker,kopp,KLH}. 

We also check that photon losses out of the cavity lead to a simple damping of the Rabi oscillations, and a faster relaxation towards the equilibrium (see Fig. 8).

\subsection{Drive Effects} 

Now, we consider the weak-coupling limit $g/\omega_0\ll 1$ which allows us to realize a pure state with a single polariton. Applying a coherent semi-classical drive to this anharmonic system can account for several non linear effects which have been explored recently \cite{bishop,schon2}. We focus on the case of a system initially prepared in its ground state and probed via an AC drive with a frequency $\omega_d$. 

In order to consider the physical effects at stake, we first consider the driven and non-dissipative Jaynes-Cummings model $(\ref{JC_drived})$.
\begin{equation}
H= \frac{\Delta}{2} \sigma^z+  \omega_0 a^{\dagger} a+ \frac{g}{2} (\sigma_{+} a+\sigma_{-} a^{\dagger})+\frac{V_0}{2} (a e^{i\omega_d t}+a^{\dagger } e^{-i\omega_d t}) .
\label{JC_drived}
\end{equation}
 We can get rid of the time-dependent part of the Hamiltonian through a unitary transformation $ | \tilde{\psi} \rangle=U(t) | \psi \rangle$ with $U(t)=\exp\left[ i \omega_d (a^{\dagger} a+\sigma_{+}\sigma_{-}) t\right]$. The evolution of $| \tilde{\psi} \rangle$ is governed by the time-independent Hamiltonian $\tilde{H}$:
\begin{equation}
\tilde{H}=\frac{\tilde{\Delta}}{2} \sigma^z+ \tilde{\omega}_0 a^{\dagger} a+ \frac{g}{2} (\sigma_{+} a+\sigma_{-} a^{\dagger})+\frac{V_0}{2} (a +a^{\dagger } ),
\label{drived_h}
\end{equation}
where $\tilde{\Delta}=\Delta-\omega_d$ and $\tilde{\omega}_0=\omega_0-\omega_d$. $\tilde{H}$ is the sum of a JC Hamiltonian with renormalized energies $\tilde{\Delta}$ and $\tilde{\omega}_0$, and a time independent driving term. It is convenient to express the last term of Eq. (\ref{drived_h}) in the dressed basis $\mathcal{B}=\{|g\rangle,|1,-\rangle,|1,+\rangle,|2,-\rangle,|2+\rangle,... \}$ of the coupled system \cite{driving_dressed_basis}

\begin{figure}[h!]
\center
\includegraphics[scale=0.4]{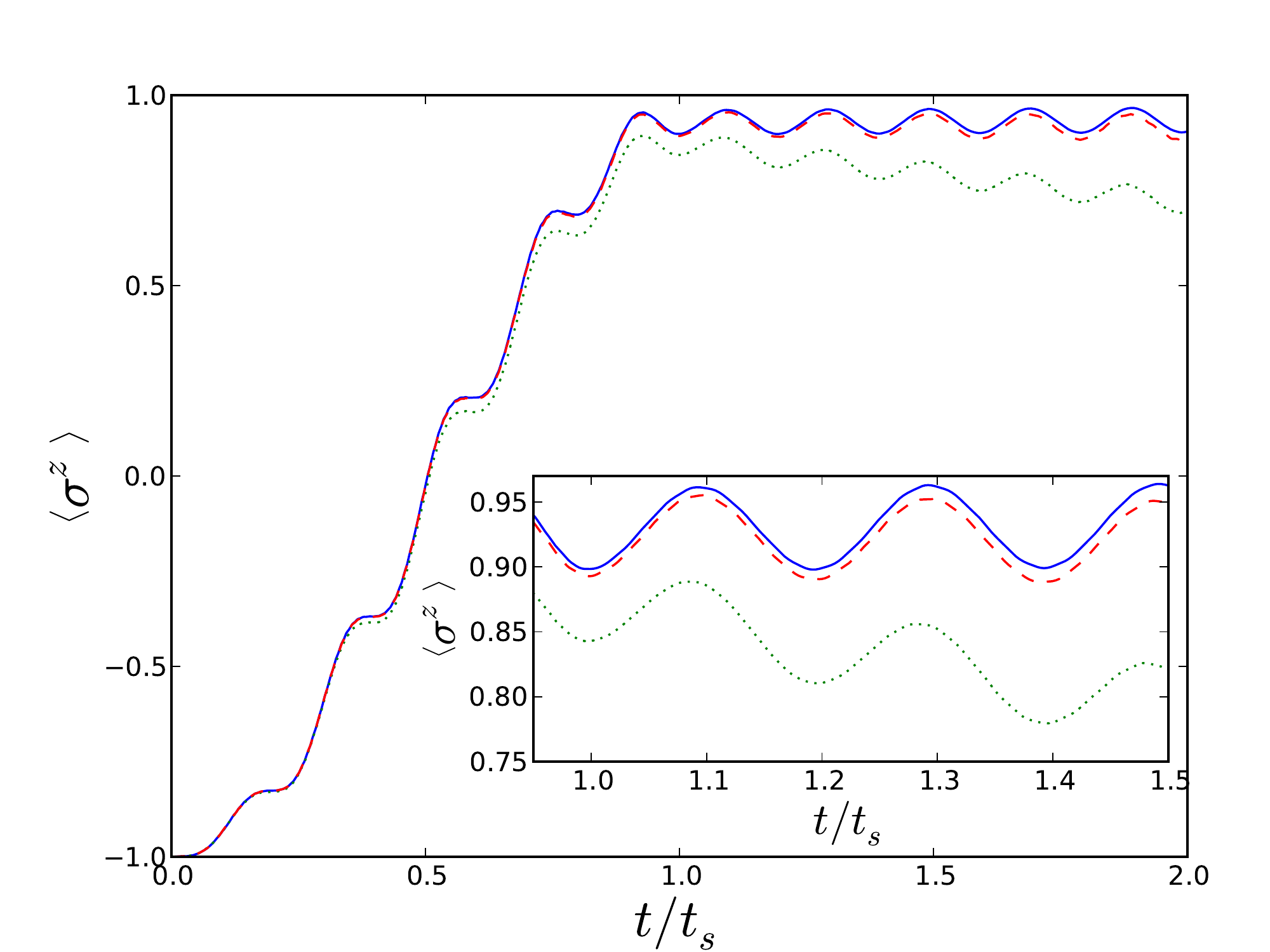}  
\caption{(Color online) Dynamics of $\langle \sigma^z \rangle$ with the system initially in its ground state; parameters are  $g/\omega_0=0.02$, $\Delta/\omega_0=0.9$ and $V_0/\omega_0=0.1$. A driving AC field is applied until the system has reached the first lower polariton; $t_s$ refers to the time at which we switch off the drive. The blue curve is the ideal dissipationless case $(\alpha=0)$ and diagonalisation in the dressed states basis gives the same result with the resolution of the figure. The dashed red curve is for  $\alpha=10^{-5}$ and the dotted green curve for $10^{-4}$; $\omega_c = 100~\omega_0$ (see also inset).}
\label{drived_sigma}
\end{figure}

\begin{align}
a +a^{\dagger }=&\ \beta_1 |1,+\rangle\langle g |+\alpha_1 |1,-\rangle\langle g | \notag \\
+\sum_{n=1}^{\infty}& \left[\sqrt{n+1} \beta_n \beta_{n+1}+\sqrt{n} \alpha_n \alpha_{n+1}\right] |n+1,+\rangle\langle n,+|  \notag \\
+& \left[\sqrt{n} \beta_n \beta_{n+1}+\sqrt{n+1} \alpha_n \alpha_{n+1}\right]         |n+1,-\rangle\langle n, -|  \notag \\
+& \left[\sqrt{n+1} \alpha_{n+1} \beta_{n}-\sqrt{n}\alpha_n \beta_{n+1}\right]     |n+1,-\rangle\langle n,+| \notag \\
 +&\left[\sqrt{n+1} \beta_{n+1} \alpha_{n}-\sqrt{n} \alpha_{n+1} \beta_{n}\right]  |n+1,+\rangle\langle n,-|   \notag \\
 +&\textrm{h.c.}.
 \label{transition_dressed_states}
\end{align}
Driving the cavity induces transition between the dressed states, and changes the number $N$ of excitations by 
$\pm1$. The Jaynes Cummings ladder is composed of two subladders of `minus' and `plus' polaritons. Eq. (\ref{transition_dressed_states}) illustrates the fact that the coupling between states of the same sub-ladder is  stronger than the coupling between states which belong to different sub-ladders. 

We then set the drive frequency $\omega_d$ to match exactly the energy difference between the ground state and the first polariton. In the limit of infinitely small drive $V_0/g\ll1$ the dynamics shows complete semi-classical Bloch oscillations of frequency $\alpha_1 V_0/2$ between these two levels; $\alpha_1= [(A -\delta_r)/2A]^{1/2}$ and $A=\sqrt{g^2+\delta^2}$. However, the switch-off time $t_s$ necessary to bring the system into the state $|1-\rangle$ is typically longer than the decoherence time.

\begin{figure}[h]
\includegraphics[scale=0.4]{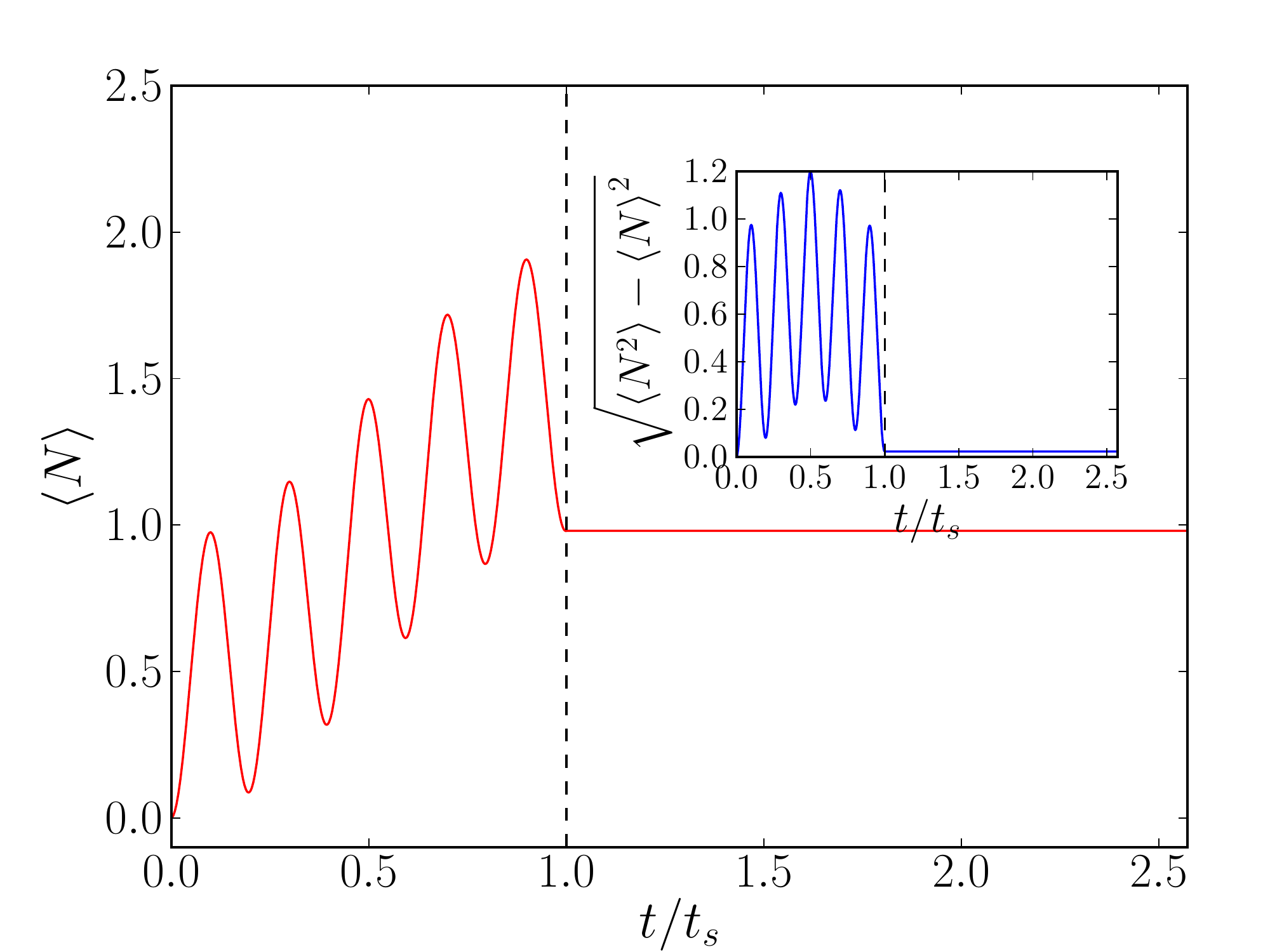}
\caption{(Color online) Dynamics of the mean number of polaritons $\langle N(t) \rangle$ in the weak-coupling $g$ limit, and without dissipation. Parameters are  $g/\omega_0=0.02$, $\Delta/\omega_0=0.9$ and $V_0/\omega_0=0.1$.  From the ground state, the system is brought into a non-trivial polaritonic final state by driving the cavity. The black dashed line refers to the moment when the AC coherent drive is switched off. Inset: Standard deviation with the same parameters.}
\label{drived_N}
\end{figure}

In the general case, we can compute the occupancies of all the levels associated with the JC ladder (see Fig. \ref{bloch_siegert}). The price to pay for an increase of the drive strength is the subtle interplay of the upper levels. But, the anharmonicity of the JC ladder makes it possible to quantitatively reach the first polariton beyond the linear response limit. The dynamics of $\langle \sigma^z\rangle$ is shown in Fig. \ref{drived_sigma}, by applying both the stochastic approach and exact diagonalisation in the dressed state basis for $\alpha=0$. The mean number of photons $\langle a^{\dagger} a \rangle$ is also evaluated , which enables us to evaluate the mean number of polaritons $\langle N\rangle = \langle a^{\dagger} a\rangle + (\langle \sigma^z\rangle+1)/2$ and the standard deviation associated to this observable (see Fig. 10). 

The order of magnitude of the time $t_s$ at which we stop the drive (with this drive setup) enables us to minimize the effect of dissipation. At weak dissipation (essentially $\alpha\lesssim 10^{-5}$), we can see that it is possible to realize temporarily an almost pure polaritonic state on one cavity.

\subsection{Lattice effects}

\begin{figure}[h!]
\center
\includegraphics[scale=0.4]{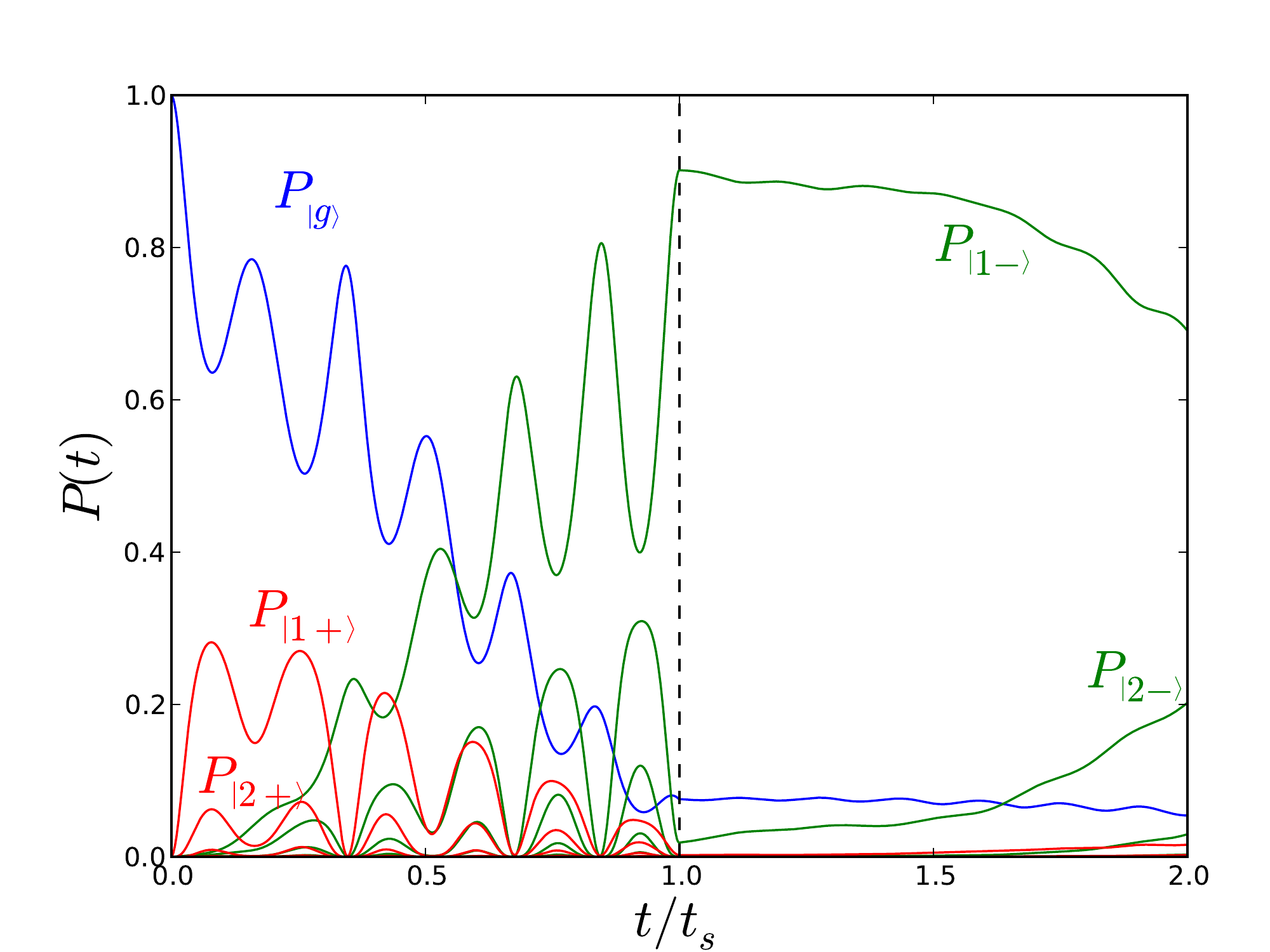}  
\caption{(Color online) Probabilities of level occupancies for the driven and dissipative case. The probability to find the system in the ground state is in blue. The probabilities corresponding to the `-' subladder are in green, while the probabilities corresponding to the `+' subladder are in red.  The system initially in its ground state; parameters are  $g/\omega_0=0.02$, $\Delta/\omega_0=0.9$ and $V_0/\omega_0=0.1$ and $\kappa/\omega_0=0.04$. A driving AC field is applied; $t_s$ refers to the time at which we switch off the drive. We remark that the green `plateau' with one polariton $|1-\rangle$ is far from $1$ and drops off rapidly. The polaritonic state is not stable. }
\label{drived_kappa}
\end{figure}

This analysis may have further implications in the realization of a driven polariton Mott state in arrays of electromagnetic resonators \cite{houck}.  Let $\kappa$ denote the (capacitive) coupling between cavities. The Hamiltonian governing the lattice system reads:

\begin{equation}
H=\sum_j H_j+ \kappa \sum_{\langle i,j \rangle} \left(a_j^{\dagger}a_i +h.c. \right),
\label{JC_drived_array}
\end{equation}
where $H_j$ is given by Eq. (\ref{JC_drived}) for each site. We then treat the photonic coupling term in a mean field manner, which results in an additional effective drive term whose strength depends on the mean field parameter $\langle a(t) \rangle$, which we consider to be independent of the spatial site $i$. We then propagate the Schr\"{o}dinger equation with this one-site effective Hamiltonian, and compute the occupancies of all the levels. If the time $t_{\kappa}\approx 1/{\kappa}$ is much greater than the switch-off time $t_s$  we can reasonably treat the drive term individually on each site (cavity). This suggests that if the transition from the ground state to the first polariton is performed in a fast manner and if dissipation effects are weak \cite{JensSebastien} 
$(\alpha \lesssim 10^{-5})$, this results in a polariton blockade for time scales smaller than $t_{\kappa}$ \cite{photon_blockade_1}. This regime seems accessible in circuit QED experiments, where the resonator frequency $\omega_0$ ranges between $1$ Ghz and $15$ Ghz. Experimental parameters could be tuned in order to have $g/\omega_0\simeq 10^{-2}$, and $\kappa/\omega_0< 10^{-3}$ for instance \cite{JensSebastien}, which implies that $t_s \ll t_{\kappa}$.

We note however that when the interaction strength between the cavities becomes more important, this protocol is no longer valid; see Fig. (\ref{drived_kappa}).

\section{Summary}

We have addressed the dynamics of the driven and dissipative quantum Rabi model. We have made quantitative predictions for the spin dynamics which can be tested experimentally. We have also shown the possibility to temporarily reach a single polariton state at short times, which constitutes a step towards the realization of a driven Mott state of polaritons in realistic conditions \cite{JensSebastien}. The stochastic approach described in the present work could be generalized to (other) hybrid systems \cite{Gurarie,hybrid1,hybrid2,hybrid3,hybrid4,hybrid5,hybrid6,Audrey}, photon lattices \cite{houck,Greentree,Angelakis,Plenio,Schmidt,JensKaryn,Fazio,Jonathan,Cristiano,biroli,JBloch,Nice,Camille,Andrew} with artificial gauge fields \cite{DalibardGerbier,IBloch,Goldman,Carusotto,circulatorQED,Hafezi,Hafezi2,Rechtsmann,Greentree2,Alex}, and fermion systems subject to time-dependent fields (potentials) \cite{Buttiker1,Buttiker,RC1,Imry,Gabelli,Feve,RC2,RC3,RC4,RC5,Prasenjit,Michele,JulieDubois,Dubois2,Christian,Bauer,Abanin,polaron,Aditya}. Other avenues could be the study of the matter-phonon coupling, dissipative spin and Kondo models.

\section{Acknowledgements}

We have benefitted from discussions with A. Browaeys, J. Cayssol, D. Est\`eve, J. Est\`eve, G. F\`eve, J. Gabelli, L. Herviou, A. Houck, J. Keeling, J. Koch, T. Lahaye, C. Mora, O. Parcollet, A. Petrescu, M. Schiro, P. Simon, Y. Sortais. The Young Investigator Group of P.P.O. received financial support from the ``Concept for the Future" of the KIT within the framework of German Excellence Initiative. This work is supported by the LABEX PALM at Paris Saclay through the project Quantum-Dyna. This work has benefitted from discussions at Harvard for the Memorial Symposium in Honor of Adilet Imambekov.

\begin{appendix}
\section{Sampling of the fields $h_{\xi}$ and $h_{\eta}$}
\label{Appendix_A}

Here, we describe how one can sample the two stochastic variables with the correlation properties given by Eqs. (\ref{heightfunctions}). We introduce variables $\tau_k=t_k/t_f$ with $t_f$ being the final time of the experiment/simulation. Hence $Q_2 (\tau)$ and $Q_1 (\tau) \theta( \tau)$ are defined on $[-1,1]$. We extend their definitions by making them $2$-periodic functions and it is then possible to expand them into Fourier series \cite{stochastic}. 

For the Rabi problem without dissipation, we define:
\begin{align}
 h_{\xi}(t_j)&=\ i \frac{g}{\omega_0} \left( s_1 \cos \omega_0  t_j +s_2 \sin \omega_0  t_j  \right)  \notag\\
 &+\frac{g}{\sqrt{8} \omega_0} \Big\{ v_1 \phi( \tau_j)+i v_2 \phi( t_j)+v_3 \phi^*(\tau_j)+i v_4 \phi^*( \tau_j)  \Big\}   \notag\\
 &+ \sum_{m=1}^{\infty}  \phi_m(\tau_j)\left(\frac{if_m^{s}}{4}\right)^{\frac{1}{2}} (u_{1,m} +i u_{2,m} ) \notag\\
&+ \sum_{m=1}^{\infty}\phi_m^*(\tau_j)   \left(\frac{if_m^{s}}{4}\right)^{\frac{1}{2}}(u_{3,m}+iu_{4,m}) ,
   \end{align}
   \begin{align}
   h_{\eta}(t_j)=&\sum_{m=1}^{\infty} \phi_m(\tau_j)\left(\frac{if_m^{s}}{4}\right)^{\frac{1}{2}} (u_{1,m} -i u_{2,m} ) \notag\\
    +& \sum_{m=1}^{\infty} \phi_m^*(\tau_j)   \left(\frac{if_m^{s}}{4}\right)^{\frac{1}{2}}(u_{3,m}-iu_{4,m}) \notag \\
    +&\frac{g}{\sqrt{8} \omega_0 } \Big\{ v_1 \phi^*(\tau_j)-i v_2 \phi^*( \tau_j)-v_3 \phi(\tau_j)+i v_4 \phi(\tau) \Big\} ,
   \label{hs}
 \end{align}
where $\phi(\tau)=\exp(i  \omega_0 \tau t_f)$, and $\phi_m(\tau)=\exp(i m \pi \tau)$. The choice of a Fourier basis is particularly well-suited for the Rabi model where coupling functions are trigonometric functions. We have an analytical expression for the Fourier coefficients $\{f_m^{s}= (g^2/\omega_0^2) \int_{-1}^1  d\tau \theta(\tau) \sin \omega_0 t_f \tau \cos m \pi \tau = (g^2 t_f/\omega_0)  \left[1-(-1)^m \cos \omega_0 t_f \right]/(\omega_0^2 t_f^2-m^2\pi^2) \}$. The convergence is then controlled until ratios $g/\omega_0$ around unity for experiment/simulation times of the order of $2\pi /\omega_0$. 

A slightly different procedure for the ohmic spin-boson model in the scaling regime ($\Delta/\omega_c \ll 1$ and $0<\alpha<1/2$) is used in Ref. \cite{stochastic}. 

\end{appendix}

\bibliographystyle{apsrev4-1}

\end{document}